\documentclass[aps,prb,fleqn,twocolumn,showpacs]{revtex4-1}
\usepackage{epsfig}
\usepackage{graphicx}
\usepackage{amsfonts,amssymb}
\usepackage{amsmath}
\usepackage{color}
\definecolor{refcolor}{rgb}{1.0,0.0,0.0} 

\newcommand{\be}{\begin{equation}}
\newcommand{\ee}{\end{equation}}   
\newcommand{\bea}{\begin{eqnarray}}
\newcommand{\eea}{\end{eqnarray}}
\newcommand{\ba}{\begin{array}}
\newcommand{\ea}{\end{array}}

\newcommand{\phrl}[1]{Phys.~Rev.~Lett. {\bf #1}}
\newcommand{\phrb}[1]{Phys.~Rev.~B {\bf #1}}

\newcommand{\q}{{\bf q}}
\renewcommand{\k}{{\bf k}}

\begin{document}
\title{Orbital-lattice coupling and orbital ordering instability in iron pnictides}
\author{Dheeraj Kumar Singh$^1$}
\email{dheeraj80@hanyang.ac.kr} 
 
\affiliation{$^1$Department of Physics, Hanyang University, 
17 Haengdang, Seongdong, Seoul 133-791, Korea}

\begin{abstract}
Orbital-ordering instability arising due to the intrapocket nesting is investigated for the tight-binding models of pnictides in the presence of orbital-lattice coupling. The incommensurate instabilities with small
momentum, which may play an important role in the nematic-ordering transition, vary from model to model besides being more favorable in comparison to the
spin-density wave instability in the absence of good interpocket nesting. We also examine the doping dependence of such instabilities. The electron-phonon 
coupling parameter required to induce them are compared with the first-principle calculations.
\end{abstract}

\maketitle
\newpage
\section{Introduction}
Iron based superconductors exhibit complex phases as a function of temperature due to the intricate 
interplay of spin, orbital and lattice degrees of freedom. They display a structural tetragonal-orthorhombic phase
transition which either precedes the collinear spin-density wave (SDW) transition\cite{nandi} or occurs simultaneously.\cite{rotter} Morever, the signatures of a nematic order in
the orthorhombic phase have been obtained as in-plane anistropy in several experiments such as angle resolved photoemission spectroscopy (ARPES),\cite{yi,shimojima} 
nuclear magnetic resonance (NMR) of spin fluctuations,\cite{fu} magnetic torque measurement\cite{kasahara} etc. As revealed by ARPES measurement, a significant 
splitting of the bands with orbital character predominantly of otherwise degenerate $d_{xz}$ and $d_{yz}$ orbitals in the tetragonal symmetry 
is observed.\cite{shimojima} The origin of this ferro-type orbital order is different from that of the ($\pi, 0$)-SDW state induced orbital ordering.\cite{daghofer1,kubo,ghosh} 
Since the nematic ordering transition is marked by the simultaneous appearance of lattice distortion, orbital order as well as 
non vanishing spin-spin correlations with time-reversal invariance, it is therefore crucial to identify experimentally the primary factor responsible for
the transition. Some progress has been made in this regard through a recent NMR experiment wherein the spin-lattice relaxation rate has been observed not
to display any change at the nematic transition, thus implying a possible key role of orbital degrees of freedom.\cite{baek}

Theoretically, the focus has been either on spin driven or orbital driven nematicity. According to the former
scenario, $Z_2$ spin-nematic order can induce the orthorhombic lattice distortion as well as ferro-orbital order
involving $d_{xz}$ and $d_{yz}$ orbitals although only with a small splitting between the orbitals.\cite{fernandes1,song,fernandes2} On the other hand, several studies have suggested 
a principle role for the orbital degree of freedom.\cite{lee_orb,chen,yanagi,kontani, kontani1} Whereas the need to include both the 
spin-lattice as well as the orbital-lattice coupling has been stressed in a Monte Carlo study within a three orbital 
spin-fermion model.\cite{liang} 

In pnictides, Fe atom lies at the center of a tetrahedron, neighboring As atoms occupy the corners, which leads to the splitting of 5-fold degenerate 
$d$ levels into two sets. $t_{2g}$ consists of degenerate $d_{xz}$, $d_{yz}$, and $d_{xy}$ levels, and $e_g$ comprises of $d_{x^2-y^2}$ and $d_{3z^2-r^2}$ levels. 
The degeneracy of $t_{2g}$ levels can be partially removed by the distortion of the tetrahedron.
Several tight-binding models\cite{raghu,daghofer,yu,graser,kuroki,ikeda,calderon} have been proposed to reproduce the Fermi surface obtained  
from first-principle calculations\cite{singh,cao,xu,haule} and ARPES measurements\cite{lu,liu1,liu2} which
includes two concentric hole pockets 
around the $\Gamma$ point and an elliptical electron pocket around the M point in the folded Brillouin zone corresponding to 2Fe/cell. The nesting between 
the hole and electron pockets leads to the collinear SDW state,\cite{brydon} while doping charge carrier results into the suppression of SDW state and the appearance of 
superconductivity as in the case of cuprates. Magnetic and transport properties within some of these models such as minimal
two-orbital model of Raghu \textit{et al}.,\cite{raghu} three-orbital model of 
Daghofer \textit{et al}.,\cite{daghofer} and five-orbital models of Graser \textit{et al}.\cite{graser} and Kuroki \textit{et al}.\cite{kuroki} have been intensively studied. However, less attention has been paid to the orbital ordering
tendency in these models especially arising due to the intrapocket nesting, which is of significant interest in the physics of pnictides particularly in the context of nematic order. 

In this paper, we explore the role of orbital-lattice coupling in the orbital-ordering instability of 
some of the tight-binding models namely two, three, and five orbital models of Raghu \textit{et al}., Daghofer \textit{et al}. and Graser \textit{et. al}., respectively. 
Such instabilities may result from the intrapocket scattering enhanced by the orbital-lattice coupling. Especially, coupling to the orthorhombic distortion can remove the
degeneracy of $d_{xz}$ and $d_{yz}$ orbitals which dominate the character of the Fermi-surface in these compounds.\cite{lv2,turner} 
%\begin{figure*}
%  \includegraphics[width=0.64\columnwidth,angle=-90]{fig1a.eps}
%  \includegraphics[width=0.6\columnwidth,angle=-90]{fig1b.eps}
%\hspace*{-5mm}
%  \includegraphics[width=0.8\columnwidth,angle=-90]{fig1c.eps}
%\hspace*{-30mm}
%  \includegraphics[width=0.8\columnwidth,angle=-90]{fig1d.eps}
%  \caption{\label{fs2}aihoeiorhweqapohrqwpeiorqweiorhio}
%\end{figure*}
%\section{Theory}
\section{Hamiltonian}
The tight-binding part of the multi-orbital Hamiltonian to describe iron pncitide is given by 
 \begin{equation}
 \mathcal{H}_{0} = \sum_{\bf k} \sum_{\mu,\nu} \sum_{\sigma}  T^{\mu \nu}({\bf k}) d_{{\bf
k}\mu \sigma}^{\dagger} d_{{\bf k} \nu \sigma},
\label{tb}
\end{equation} 
where $d_{{\bf k}\mu \sigma}^\dagger$ creates an electron 
in the $\mu$-th orbital with momentum ${\bf k}$ and spin $\sigma$. $T^{\mu \nu}({\bf k)}$ are 
the hopping elements from orbital $\mu$ to $\nu$,\cite{raghu,daghofer,graser} where $\mu$ and $\nu$ belong to 
the set of five $d$-orbitals $d_{xz}$, $d_{yz}$, $d_{xy}$, $d_{x^2-y^2}$, and $d_{3z^2-r^2}$ depending on 
the model.

The interaction part is given by  
\begin{eqnarray}
\mathcal{H}_{el-el} &=& U \sum_{{\bf i},\mu} n_{{\bf i}\mu \uparrow} n_{{\bf i}\mu \downarrow} + (U' -
\frac{J}{2}) \sum_{{\bf i}, \mu<\nu} n_{{\bf i} \mu} n_{{\bf i} \nu} \nonumber \\ 
&-& 2 J \sum_{{\bf i}, \mu<\nu} {\bf{S_{{\bf i} \mu}}} \cdot {\bf{S_{{\bf i} \nu}}} + J \sum_{{\bf i}, \mu<\nu, \sigma} 
d_{{\bf i} \mu \sigma}^{\dagger}d_{{\bf i} \mu \bar{\sigma}}^{\dagger}d_{{\bf i} \nu \bar{\sigma}}
d_{{\bf i} \nu \sigma}, 
\label{int}
\end{eqnarray}
which includes the intraorbital (interorbital) Coulomb interaction term as the first (second) term. The third term describes the Hund’s coupling, and the fourth term represents 
the pair hopping energy. Rotation-invariant interaction is ensured provided that $U$ = $U^{\prime}$ + $2J$. 

Finally, we consider the orbital-lattice coupling given by 
\begin{equation}
\mathcal{H}_{e-ph} = 
\sum_{{\bf i}}
 g\mathcal{\epsilon}_{{\bf i}}\mathcal{O}_{{\bf i}}
 +\mathcal{K} {\epsilon}^2_{{\bf i}}/2,
\label{eph}
 \end{equation}
where the first term denotes the coupling of orbital degree of freedom to the orthorhombic strain (${\epsilon}_{\bf i}$) and the second term represents the potential energy due to the strain. Here, 
$\mathcal{O}_{{\bf i}}$ is the quadrupole operator or the orbital operator $(n_{{\bf i}1}-n_{{\bf i}2})$ having $B_{1g}$ symmetry. Subscript 1 and 2 have been used for $d_{xz}$ and $d_{yz}$ orbitals, respectively. 
Carrying out the quantization of the strain, $\mathcal{H}_{e-ph}$ can be written as 
$ \sum_{{\bf i}} g^{\prime} (a_{{\bf i}}+a^{\dag}_{{\bf i}})\mathcal{O}_{{\bf i}}+\sum_{{\bf i}} \omega^{\prime}(a^{\dag}_{{\bf i}}a_{{\bf i}}+1/2)$,
where $a^{\dag}_{{\bf i}}$ is the phonon creation operator, $g^{\prime} = g/\sqrt{2 \omega^{\prime} m}$, and $\omega^{\prime} = \sqrt{\mathcal{K}/m}$.
 
\section{Orbital ordering instabilities}
To investigate the orbital-ordering instability, we consider the orbital susceptibility defined as follows:
\begin{equation}
\chi^{o}(\q,i\Omega_n)= \int^{\beta}_0{d\zeta e^{i \Omega_{n}\zeta}\langle T_\zeta [{\cal O}_\q(\zeta) {\cal O}_{-\q}(0)]\rangle}.
\end{equation}
Here, $\langle...\rangle$ denotes thermal average, $T_\zeta$ imaginary time ordering, and $\Omega_n$ are the Bosonic
Matsubara frequencies. ${\cal O}_{\bf q}$ is 
obtained as the Fourier transformation of 
$\mathcal{O}_{{\bf i}}$ as defined in the previous section.

Orbital susceptibility ${\chi}^{\rm o}({\bf q})$ is calculated within the RPA-level using \cite{takimoto} 
\begin{eqnarray}
    && \hat{\chi}^{o}({\bf q}) \!=\!  \hat{\chi}({\bf q})
  [\hat{1}+\hat{U}^{\rm o}\hat{\chi}({\bf q})]^{-1},
\label{orb}
  \end{eqnarray}
where $\hat{1}$ is the $n^2 \times n^2$ unit matrix, where $n$ is the number of orbitals in the model. 
The elements of $n^2 \times n^2$ matrix $\hat{\chi}({\bf q})$ are defined by
\bea
\chi^{}_{\mu \nu, \alpha \beta}(\q, i \Omega_n)  &=& -\frac{1}{N}\sum_{\k } \sum_{i,j} 
a^{\mu^*}_{j\k + \q} a^{\nu }_{i \k} a^{\beta^* }_{i \k}  
a^{\alpha }_{j
\k + \q} \nonumber\\ &\times& \frac { n(E_{i}(\k))-n(E_{j}(\k+\q)) }{i \Omega_n+E_{i}(\k)-E_{j}(\k+\q) } .
\eea
The interaction matrix is $\hat{U}^{o} = \hat{C} + 2\hat{P}$ with nonvanishing matrix elements given as $C_{n_1 n_2, n_3 n_4} = U,\,\, -U^{\prime} + 2J,\,\, 2U^{\prime} - J$, and $J$ for 
$n_1 = n_2 = n_3 = n_4$, $n_1 = n_3 \ne n_2 = n_4$, $n_1 = n_2 \ne n_3 = n_4$, and $n_1 = n_4 \ne n_2 = n_3$, respectively. For the phonon-mediated interaction matrix $\hat{P}$,\cite{yanagi,kontani}
$P_{11,11}  =  P_{22,22} = - g^{\prime 2} D(i\Omega_m))$ and $ P_{11,22} = P_{22,11} = g^{\prime 2} D(i\Omega_m))$, where 
\begin{equation}
 D(i\Omega_m)=\frac{2\omega^{\prime}}{\Omega^{2}_{m}+\omega^{\prime 2}}.
\end{equation}
In the following, we set $J = U/6$ as suggested by the first-principle calculation.\cite{miyake} A dimensional-less electron-phonon coupling parameter is 
defined as $\tilde{\lambda}$ = $\rho \lambda$, where $\rho$ is the density of states at the Fermi level and $\lambda = g^{\prime 2} D(0)$. The matrix elements 
$\chi_{1111}(\q)$, $\chi_{2222}(\q)$, and $\chi_{1122}(\q)$ are of our interest as they contribute to the orbital susceptibility corresponding to the
operator $\mathcal{O}$ defined earlier. We also note that $\chi_{1111}(q_x, q_y)$ = $\chi_{2222}(q_y, q_x)$ due to the symmetry consideration.
\subsection{Two-orbital model of Raghu \textit{et al}.}
\begin{figure*}
  \includegraphics[width=0.64\columnwidth,angle=-90]{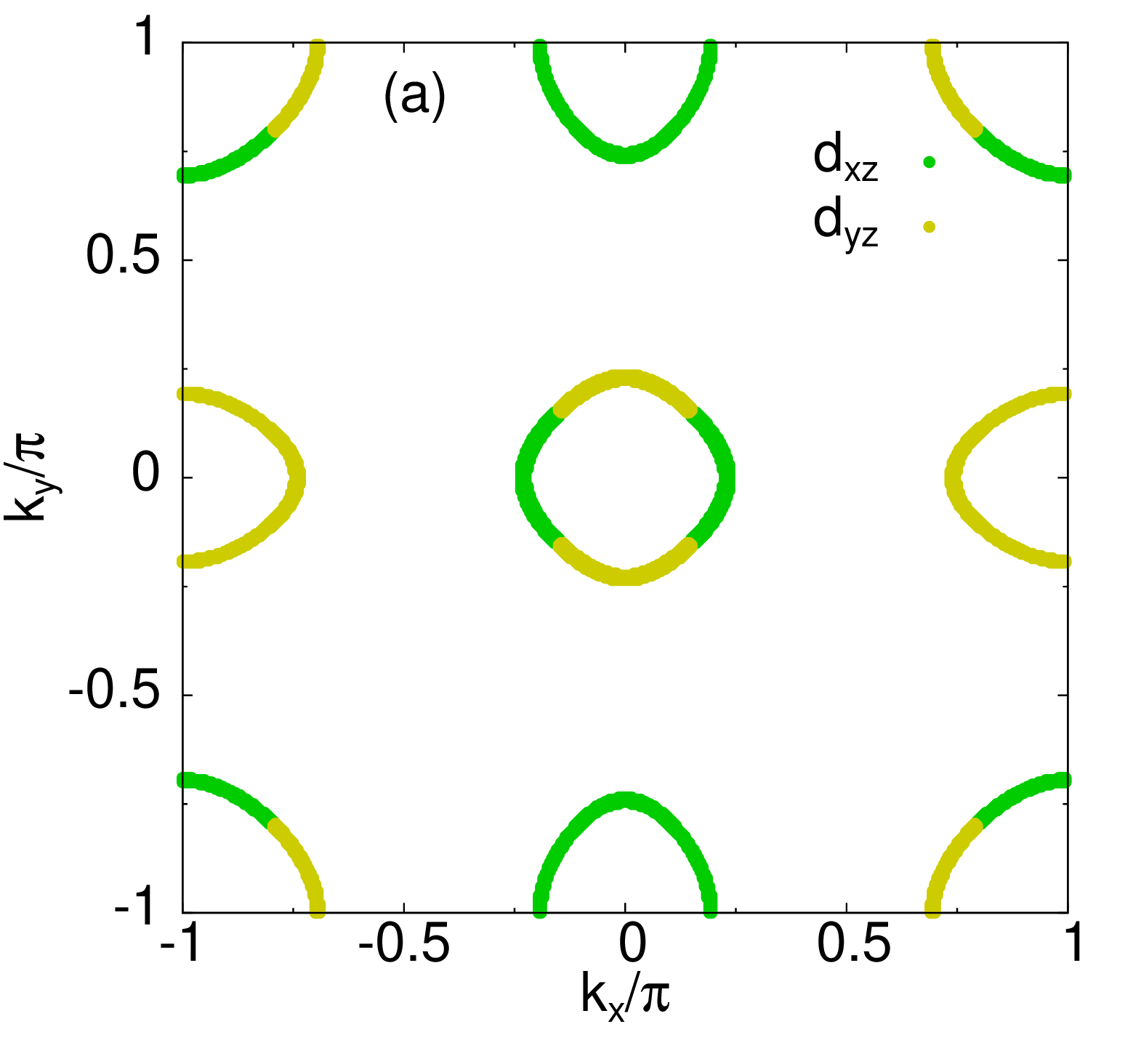}
  \includegraphics[width=0.6\columnwidth,angle=-90]{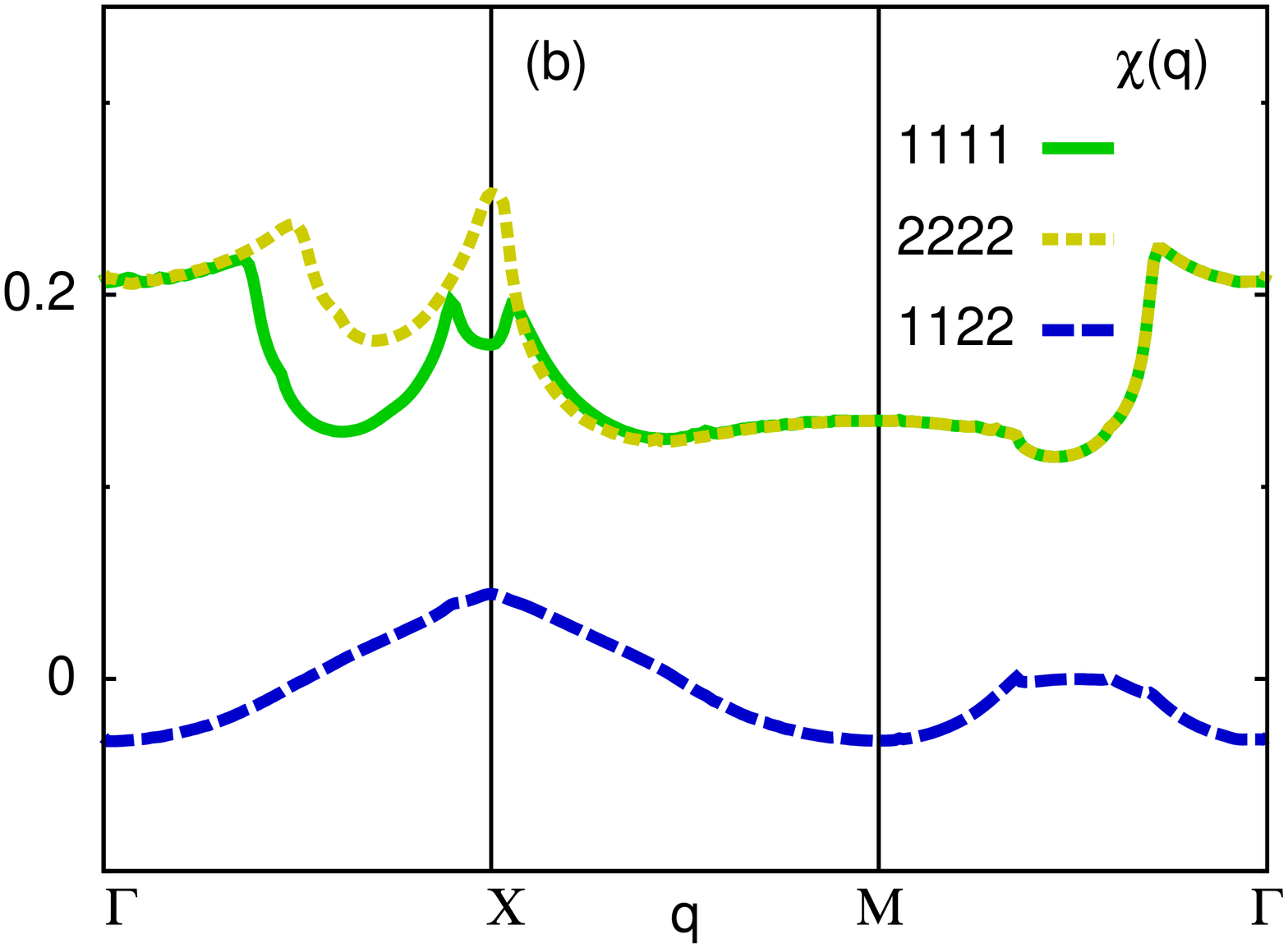}
  \includegraphics[width=0.6\columnwidth,angle=-90]{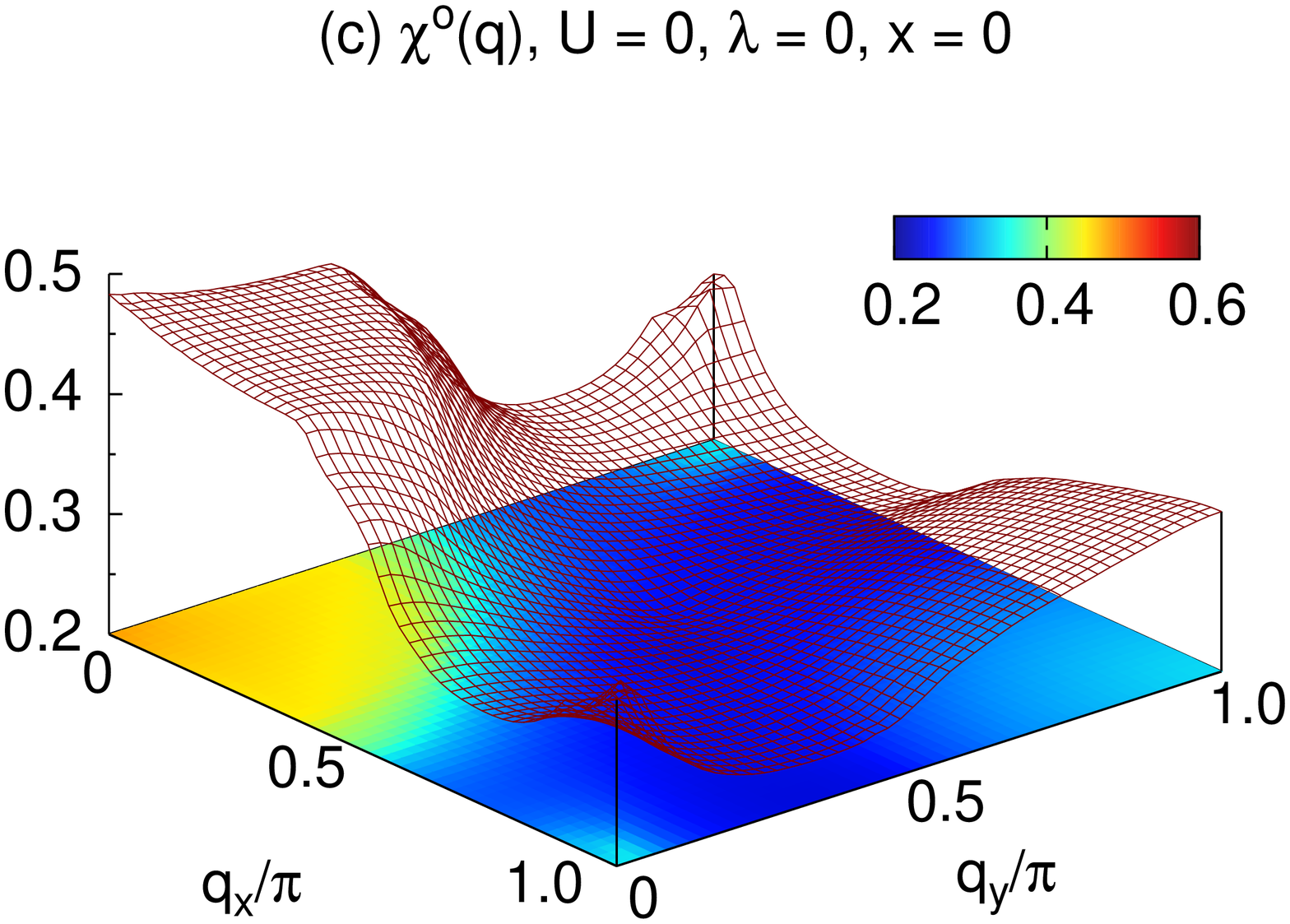}
\hspace*{-20mm}
  \includegraphics[width=0.6\columnwidth,angle=-90]{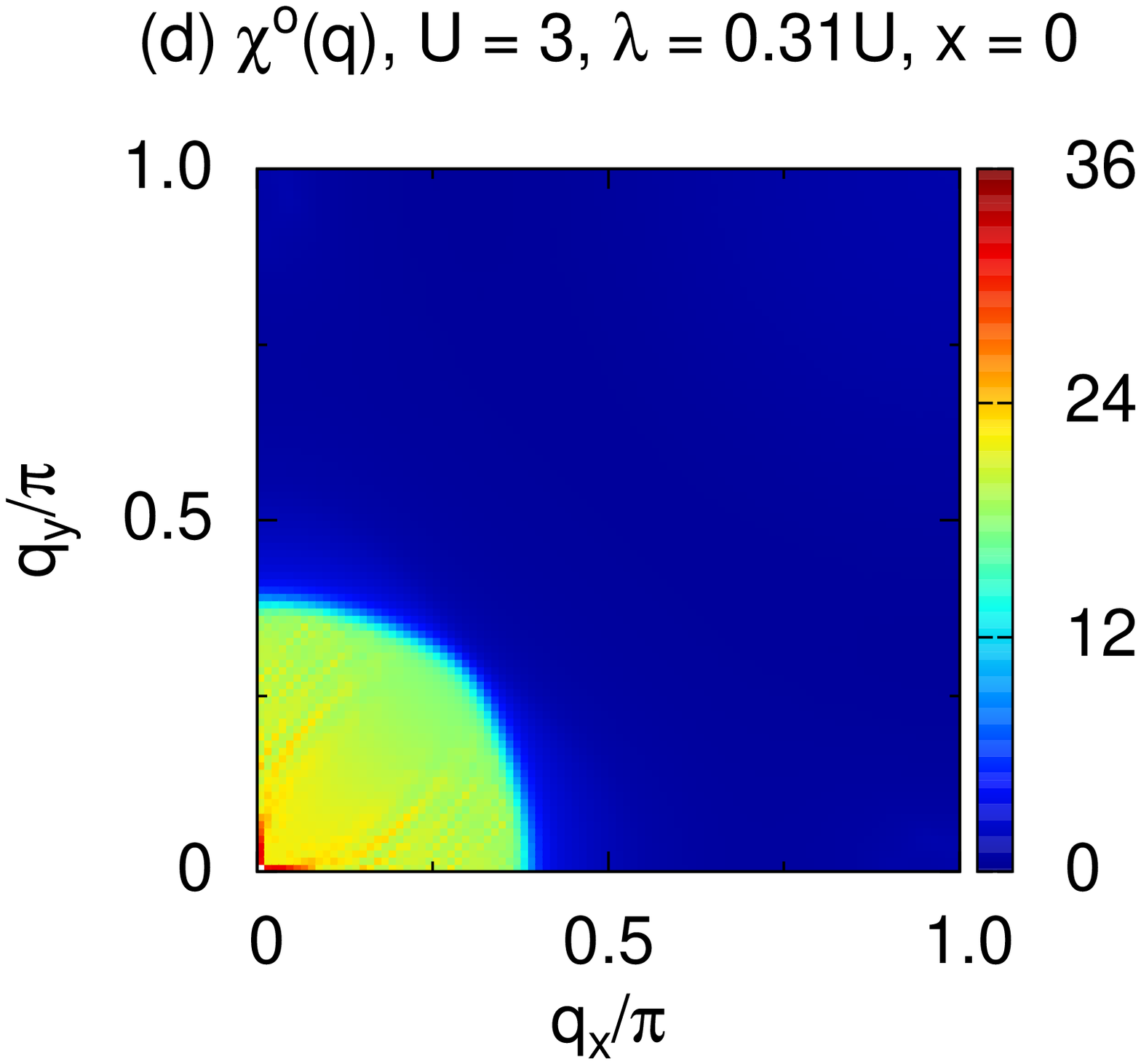}
\hspace*{-27mm}
  \includegraphics[width=0.6\columnwidth,angle=-90]{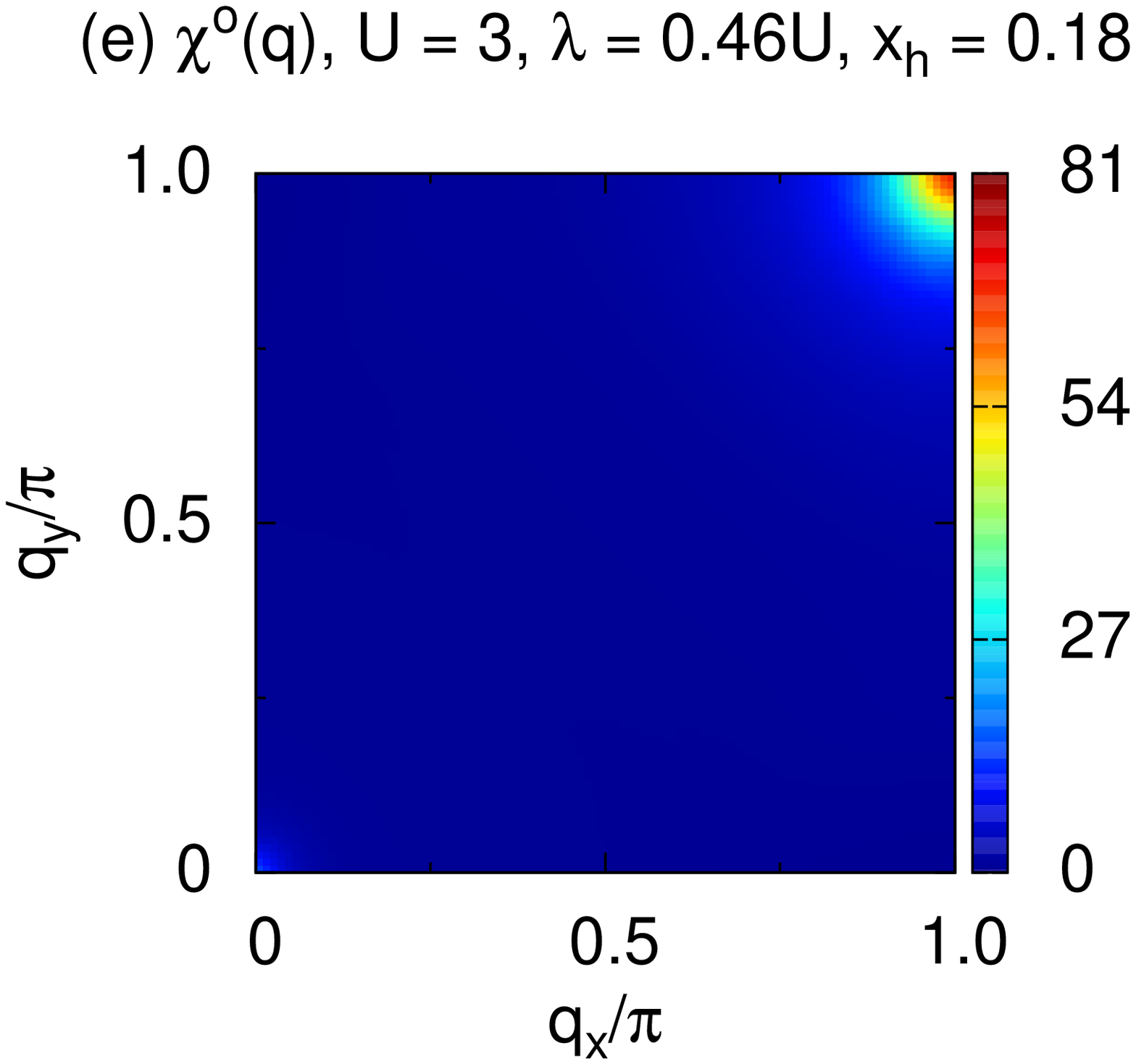}
    \caption{(a) Fermi surface for $n \approx 2$ in the two-orbital model of Raghu \textit{et al}. with predominant orbital distributions. (b) Components of 
    one bubble susceptibility contributing to the orbital susceptibility. (c) Bare orbital susceptibility for undoped case. RPA-level orbital susceptibility for (d)
    $x_h$ = 0, $\lambda = 0.31U$ and (e) $x_h$ = 0.18, $\lambda_{\q_1} = 0.46U$, where  $U = 3.0$.}
\label{raghu}
  \end{figure*}
\begin{figure*}
  \includegraphics[width=0.64\columnwidth,angle=-90]{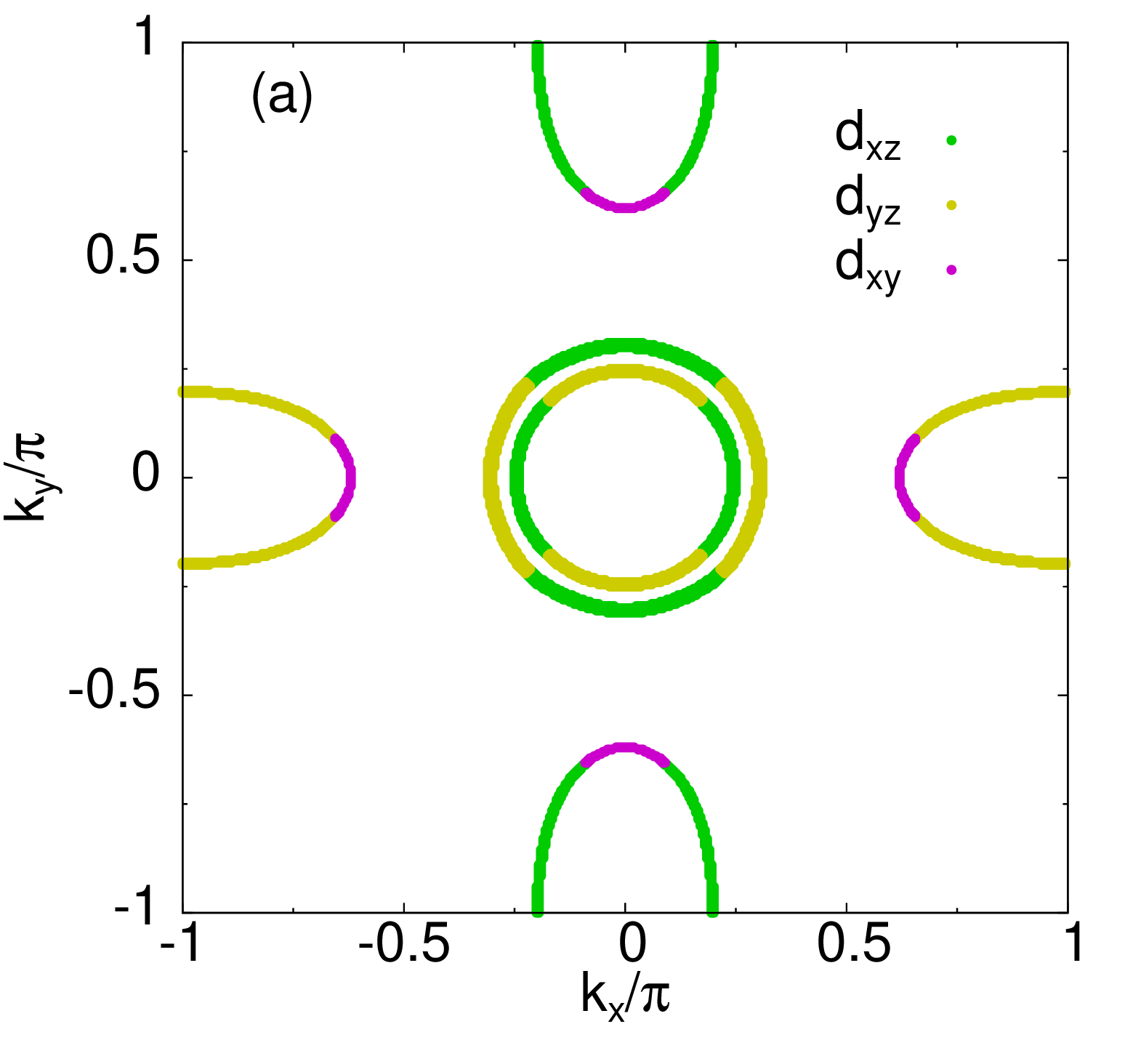}
  \includegraphics[width=0.6\columnwidth,angle=-90]{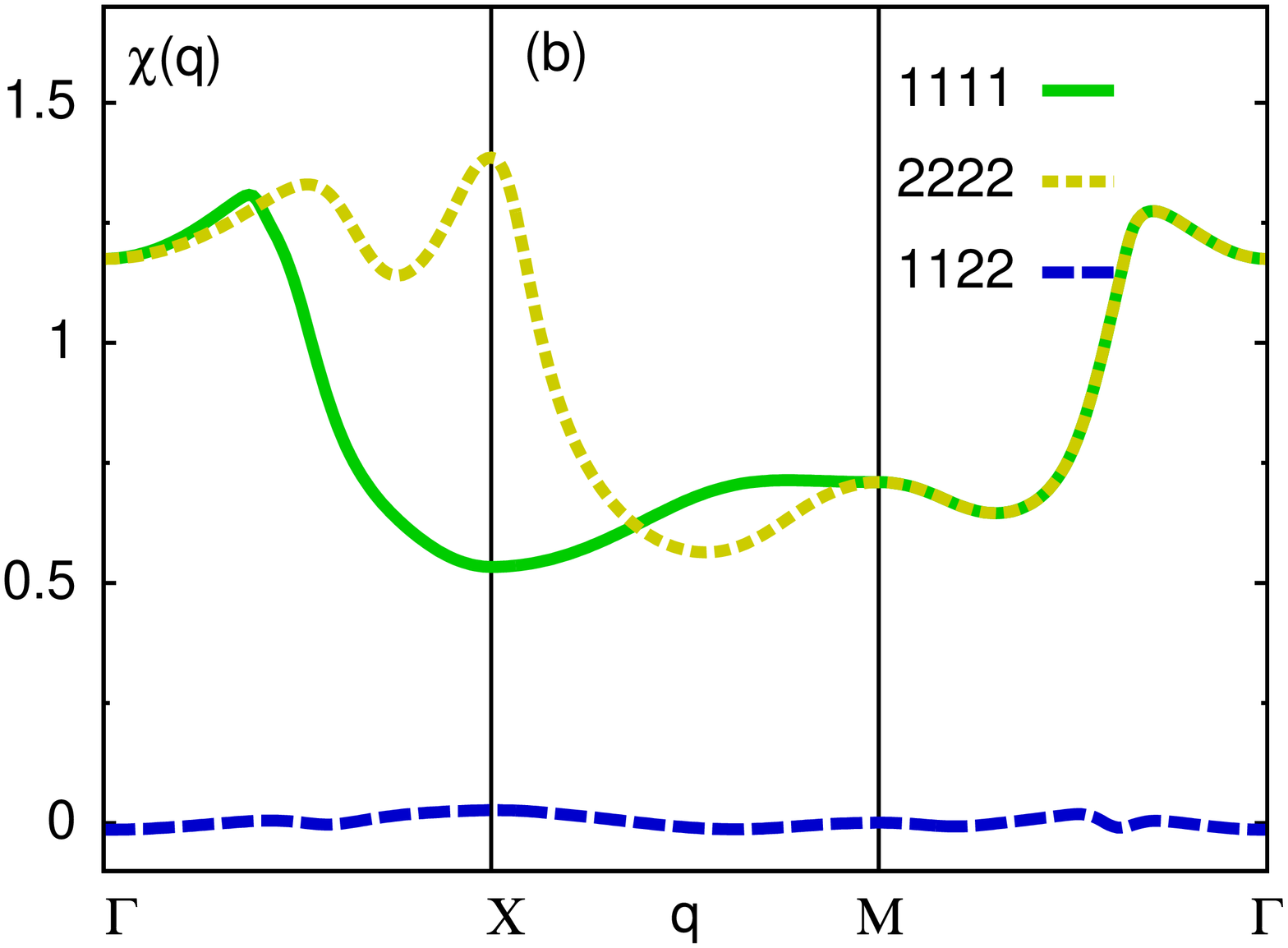}
  \includegraphics[width=0.6\columnwidth,angle=-90]{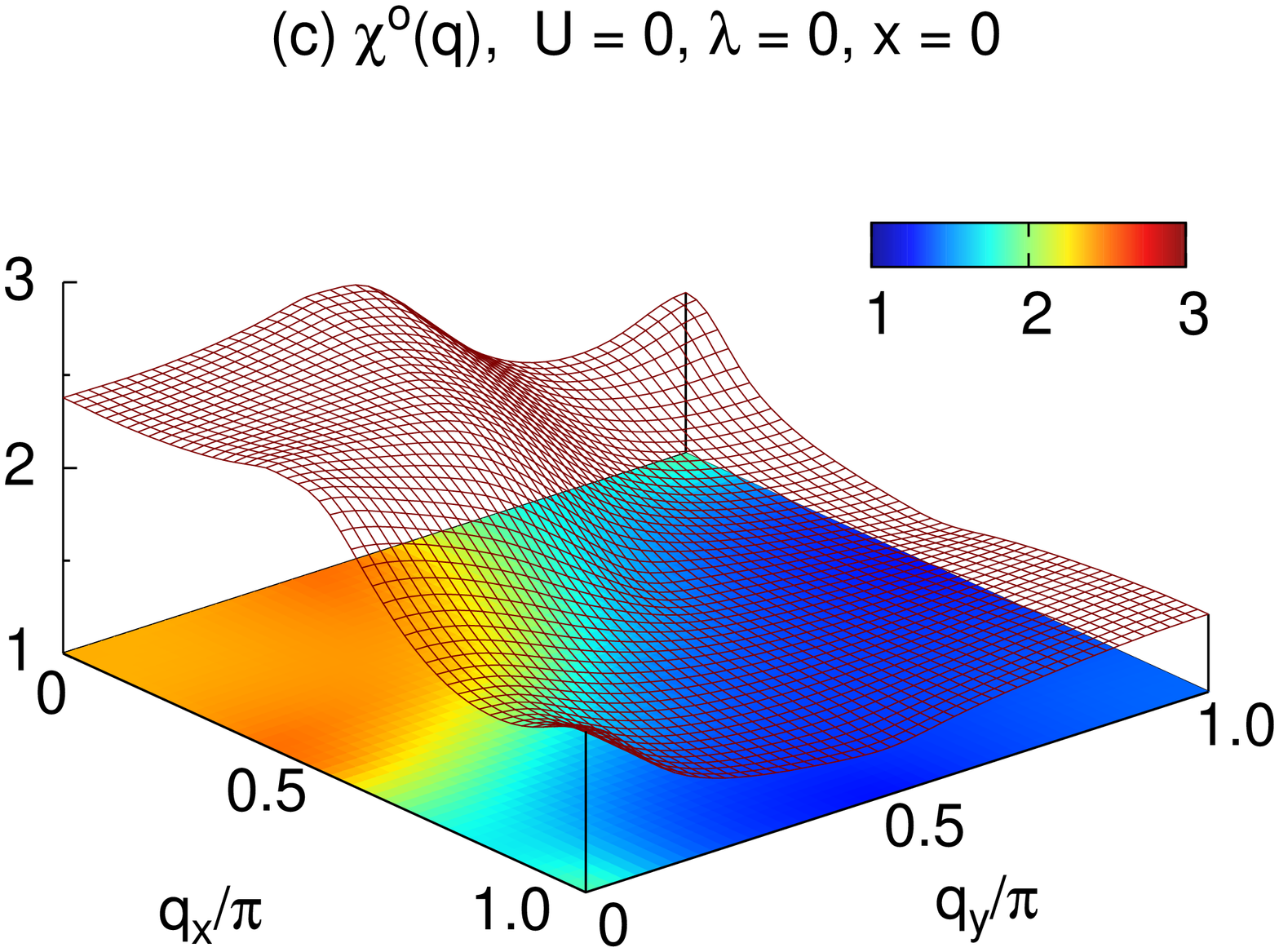}
\hspace*{-20mm}
  \includegraphics[width=0.6\columnwidth,angle=-90]{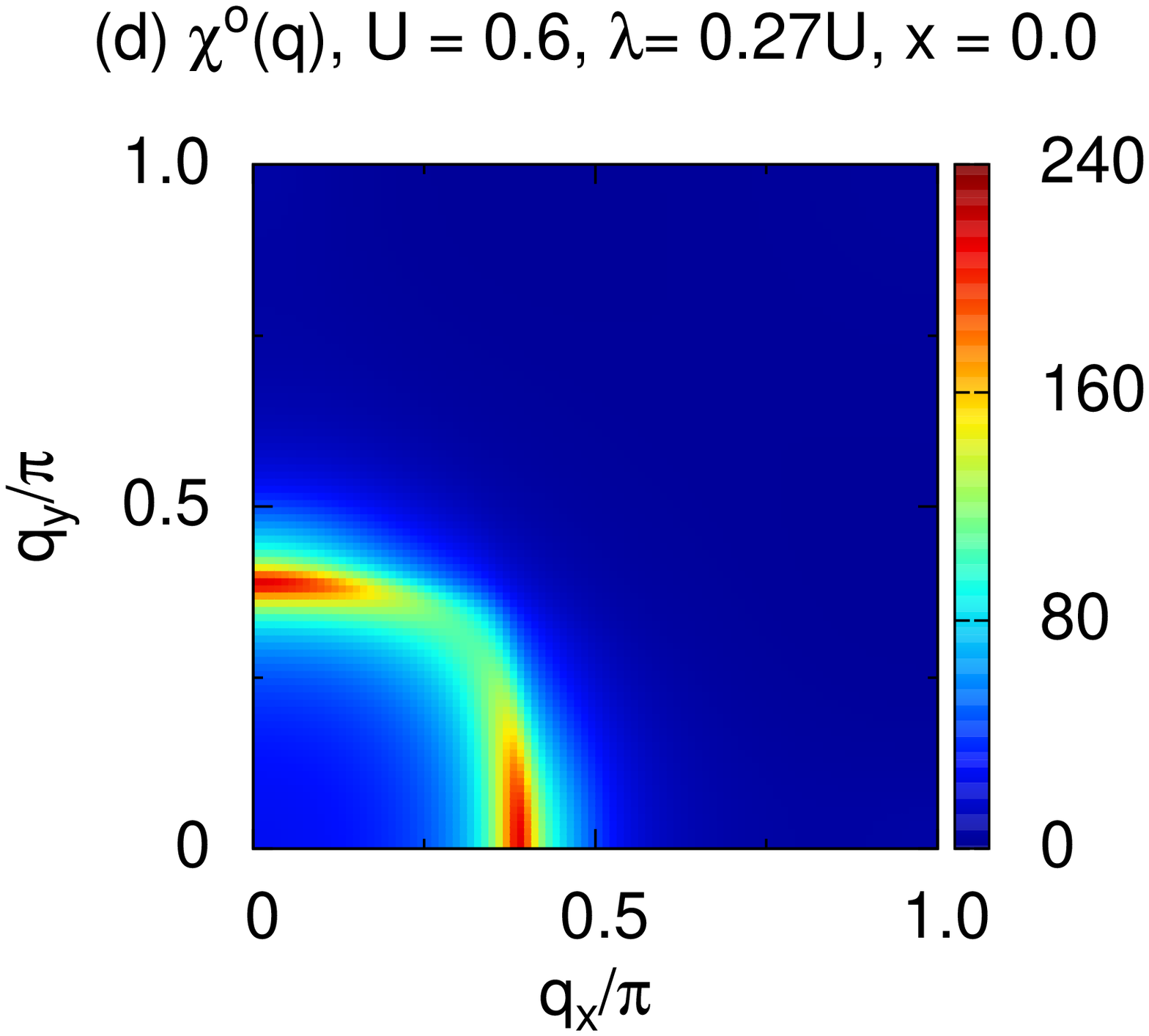}
\hspace*{-27mm}
  \includegraphics[width=0.6\columnwidth,angle=-90]{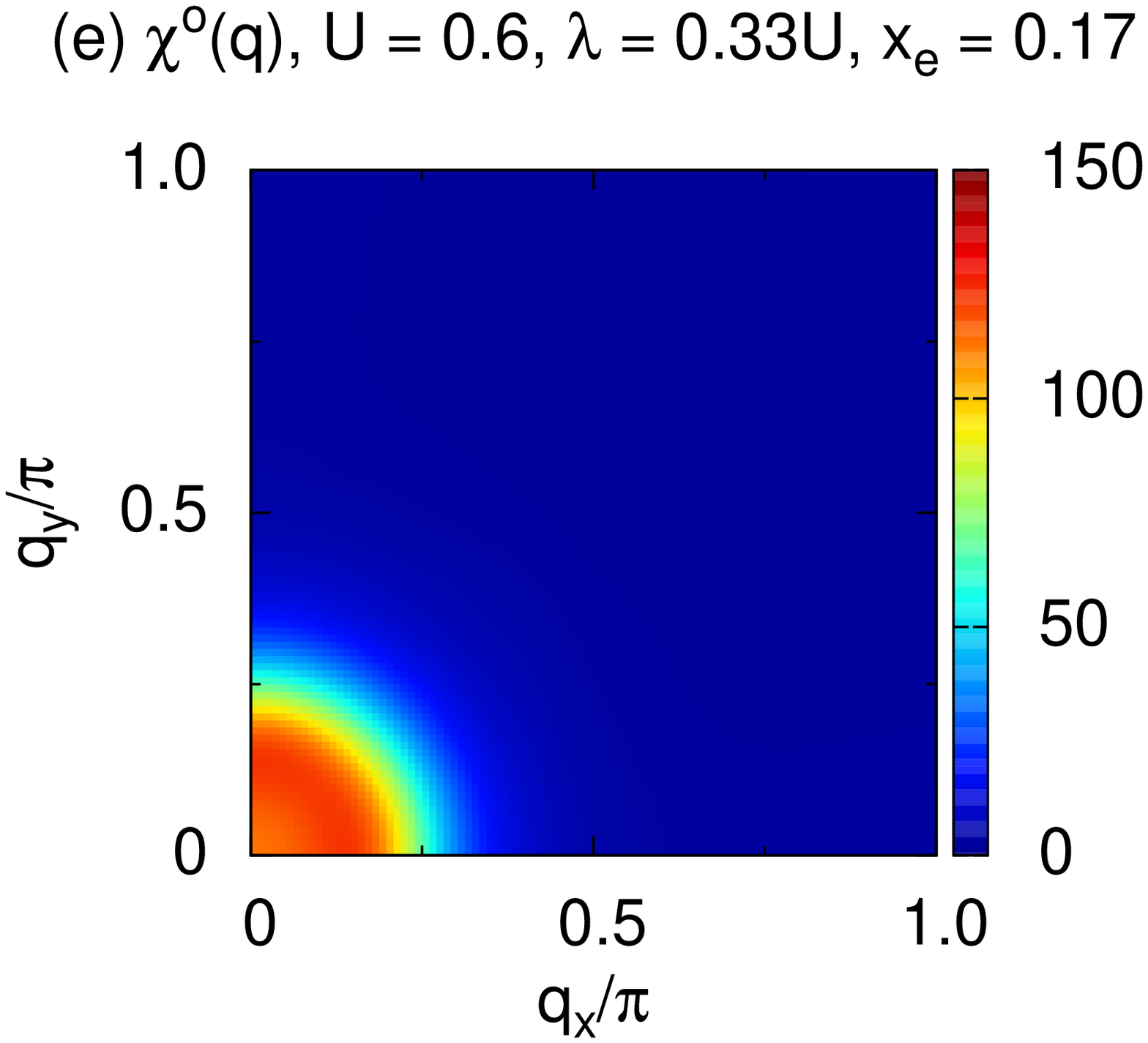}
    \caption{(a) Fermi surface for $n \approx 4$ with the orbital character in the three-orbital model of Daghofer \textit{et al}.. (b) Elements of 
    one bubble susceptibility. (c) Bare orbital susceptibility for zero doping. RPA-level orbital susceptibility for 
    (d) $x_e$ = 0, $\lambda = 0.27U$ and (e) $x_e$ = 0.17, $\lambda_{} = 0.33U$, where $U = 0.6eV$.}
\label{daghofer}
  \end{figure*}
\begin{figure*}
  \includegraphics[width=0.64\columnwidth,angle=-90]{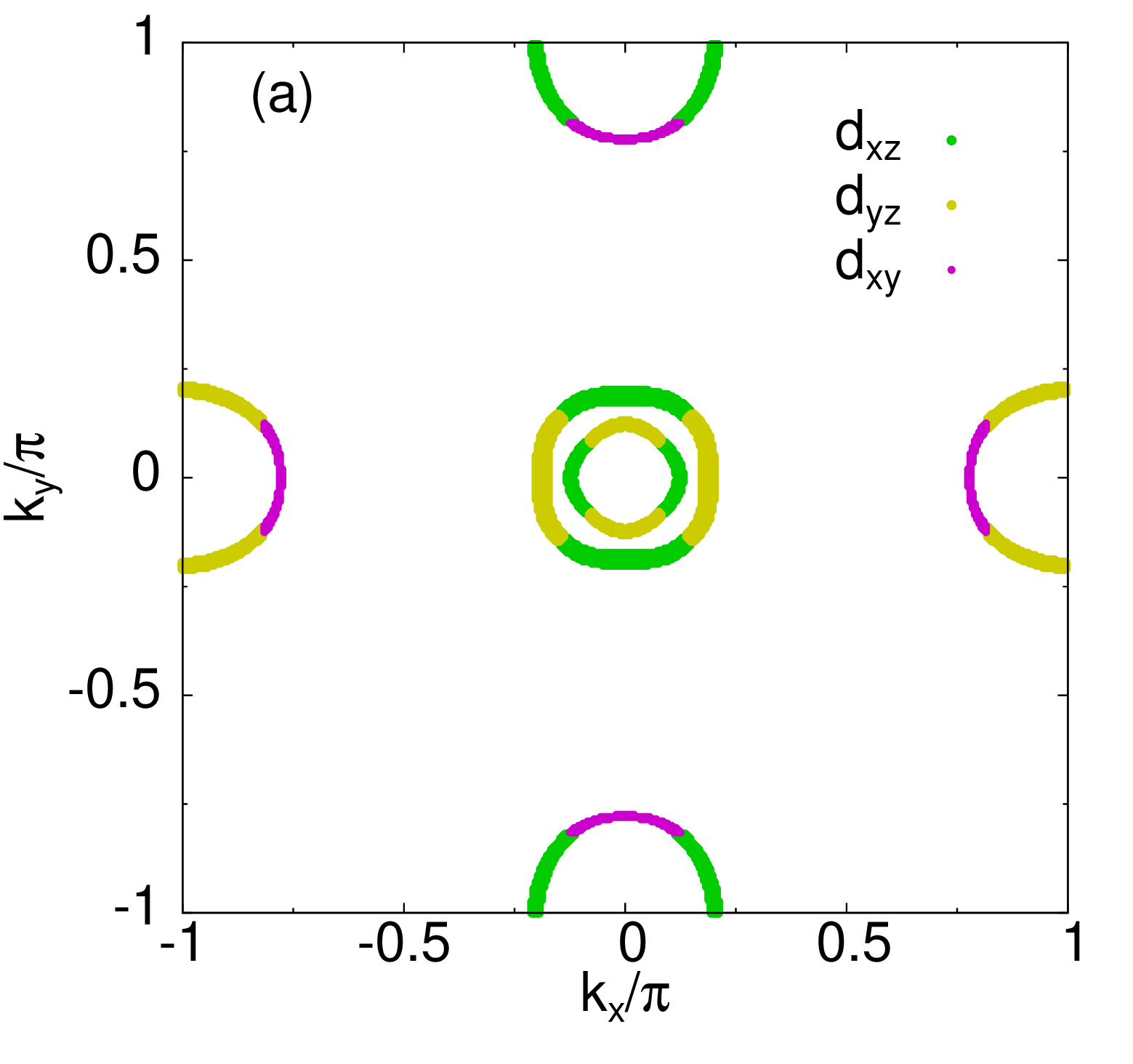}
  \includegraphics[width=0.6\columnwidth,angle=-90]{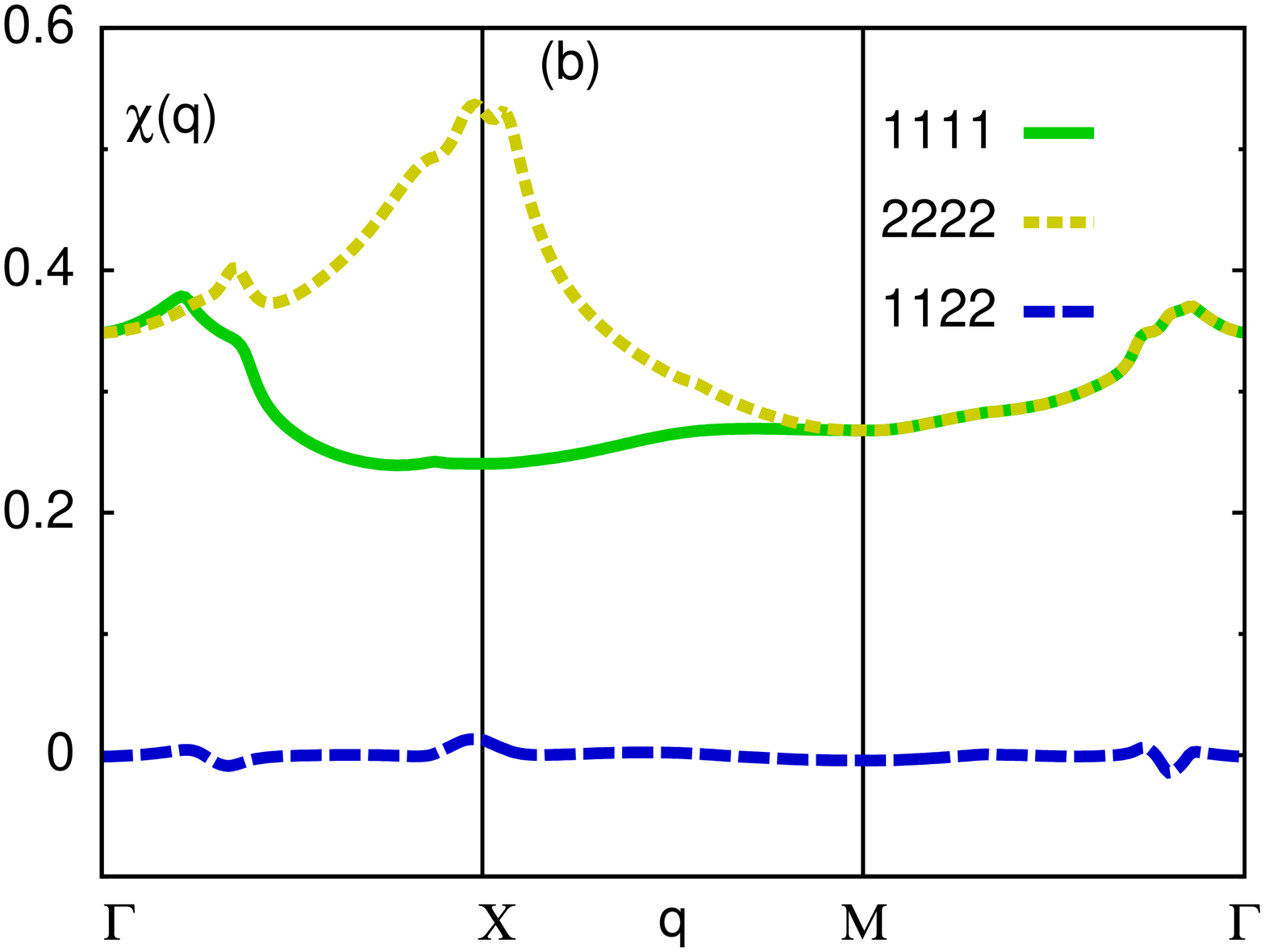}
  \includegraphics[width=0.6\columnwidth,angle=-90]{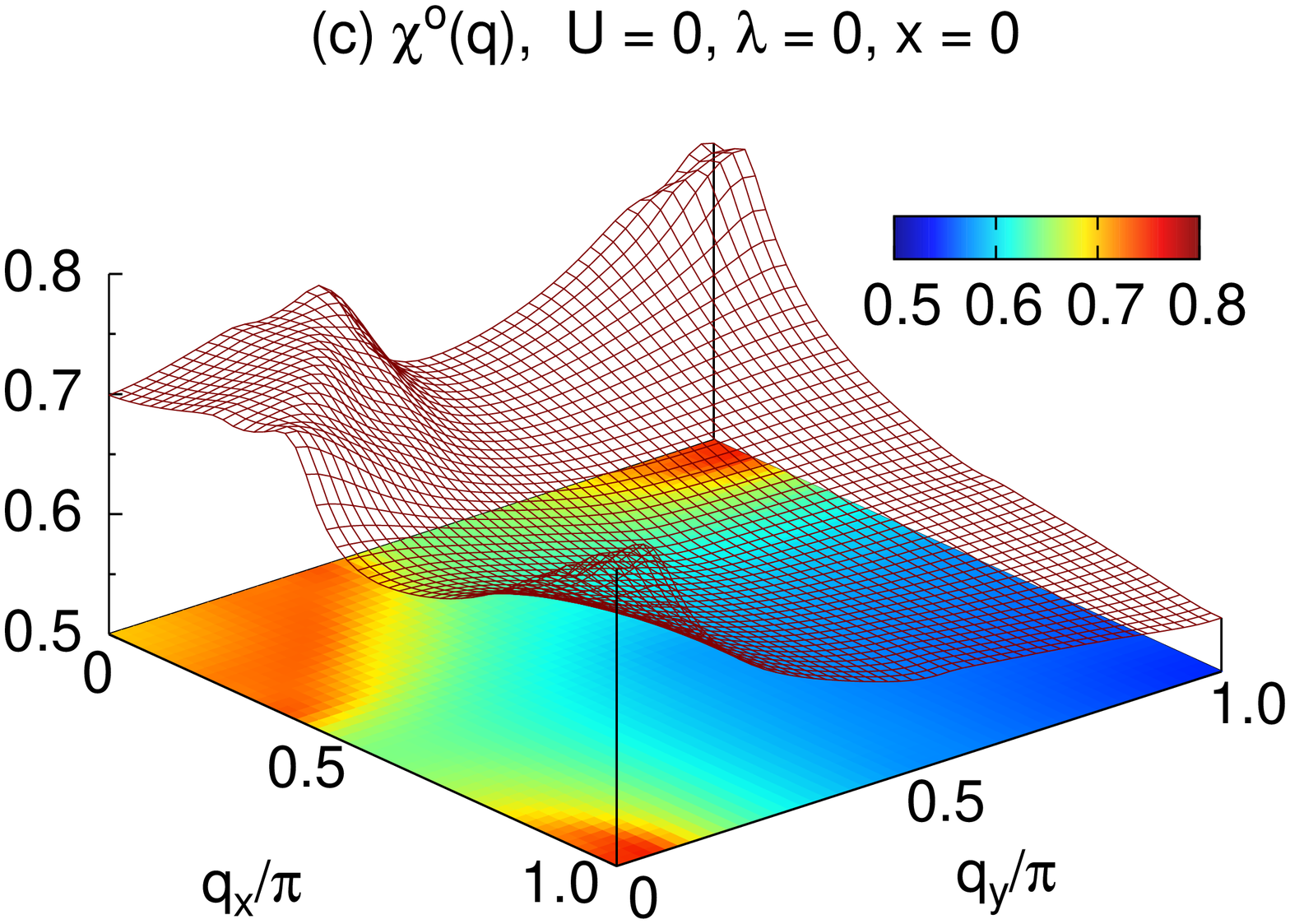}
\hspace*{-20mm}
  \includegraphics[width=0.6\columnwidth,angle=-90]{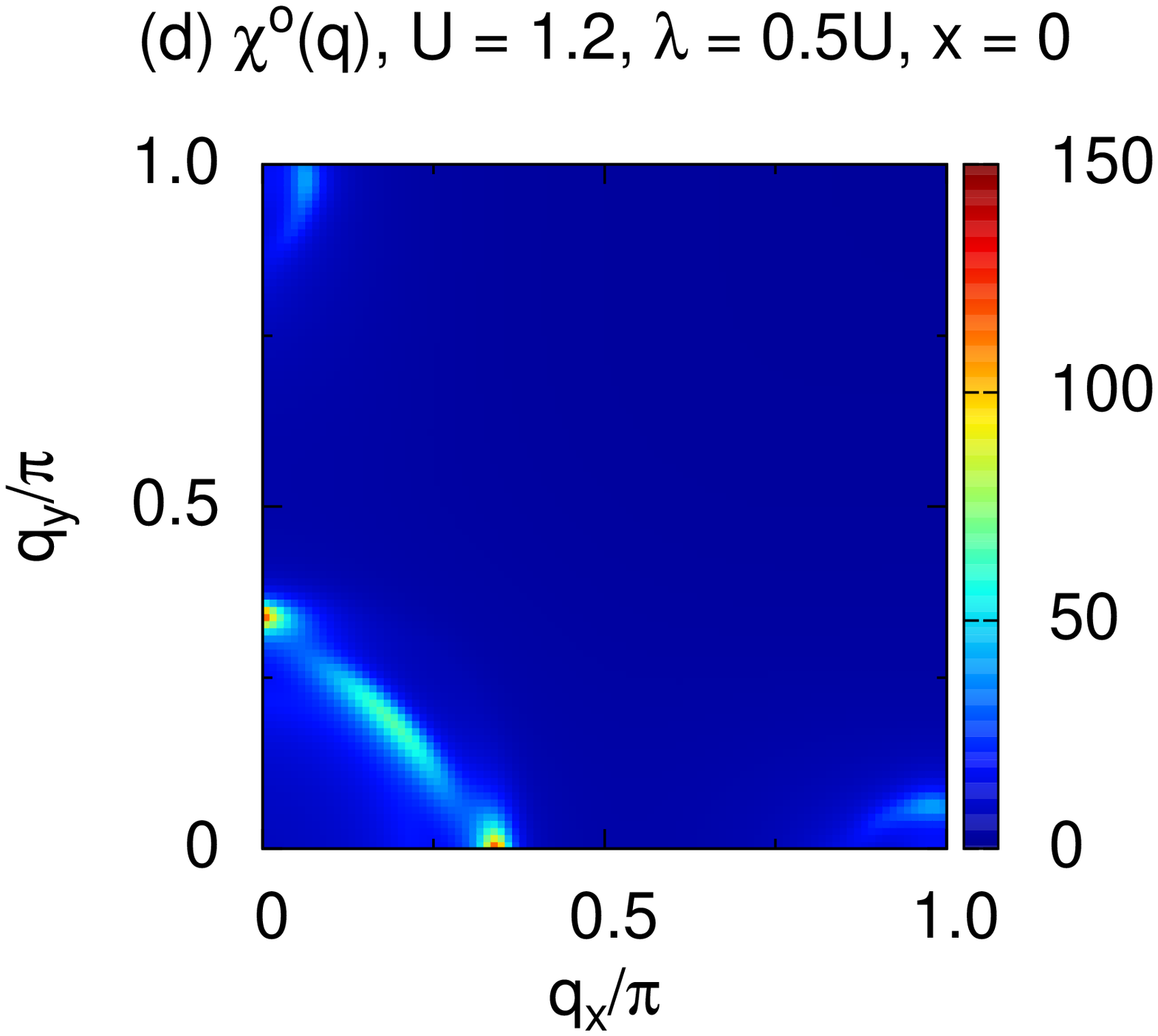}
\hspace*{-27mm}
  \includegraphics[width=0.6\columnwidth,angle=-90]{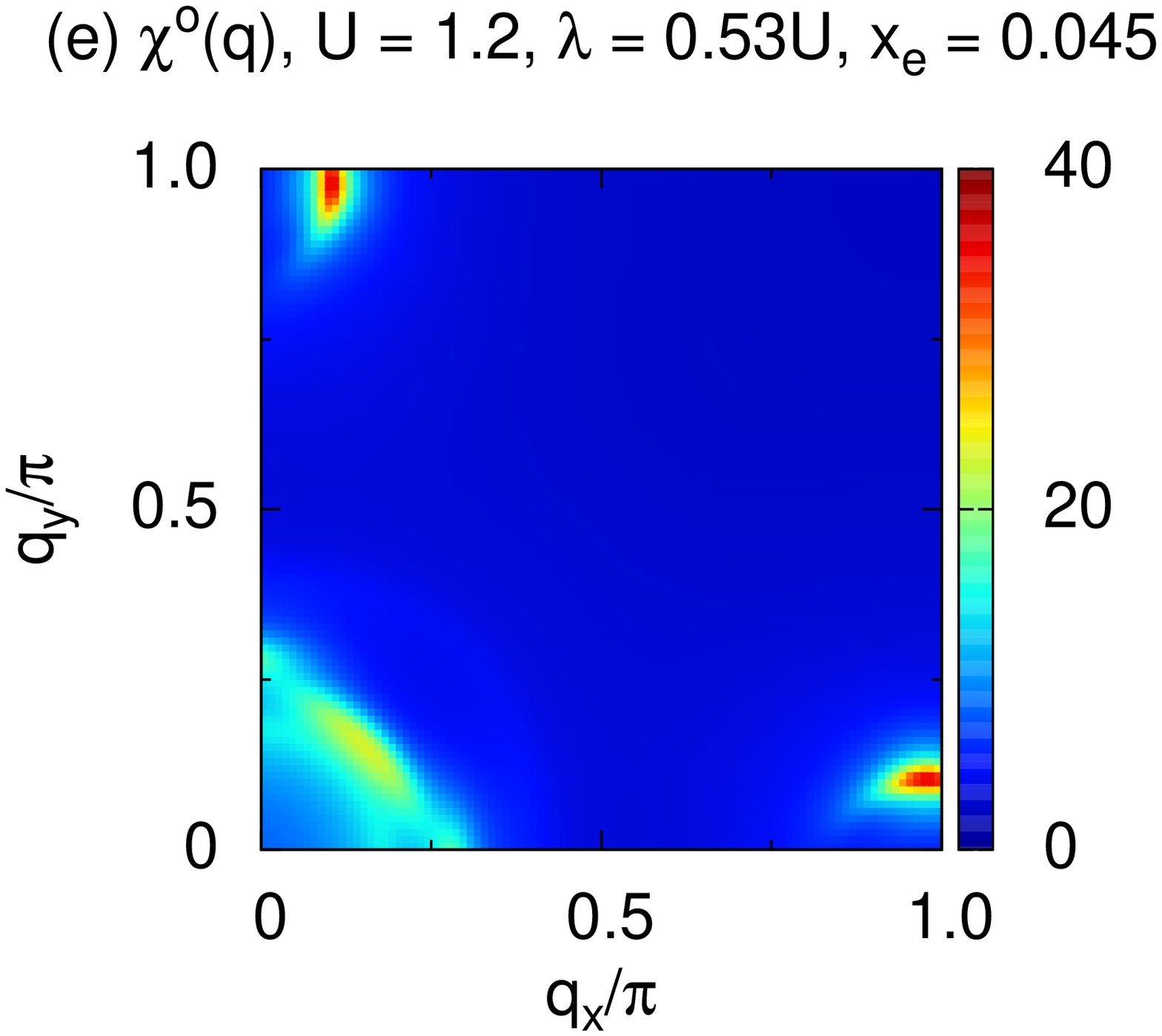}
    \caption{(a) Fermi surface for $n \approx 6$ in the five-orbital model of Graser \textit{et al}. with predominant orbital distributions. (b) Relevant components of 
    one bubble susceptibility. (c) Bare orbital susceptibility in the undoped case. RPA-level orbital susceptibility for 
    (d) $x_e$ = 0, $\lambda = 0.5U$ and (e) $x_e$ = 0.045, $\lambda = 0.53U$, where $U = 1.2eV$.}
\label{graser}
  \end{figure*}
  
Fig. 1(a) shows the Fermi surface with predominant orbital character for $\mu = 1.45$ ($n \approx$ 2) in the two orbital model of Raghu \textit{et al}. with basis consisting of $d_{xz}$ and $d_{yz}$ orbitals. It consists of one hole and one electron pocket at (0, 0) and ($\pi, 0$), respectively. An 
additional hole pocket is obtained around ($\pi, \pi$) instead of (0, 0).
An important shortcoming of this minimal model is the exclusion of $d_{xy}$ orbital, which can contribute
significantly to the Fermi surface as reported by the band-structure calculations. Despite these limitations, this model 
is one of the simplest models, which has been widely employed to study various properties including the $(\pi, 0)$-SDW state in the iron pnictides.   

The components of bare susceptibility corresponding to the Fermi surface (Fig. 1(a)) are shown in Fig. 1(b). Both 
$\chi_{1111}(\q)$ and $\chi_{2222}(\q)$ exhibit peaks near $(0.3\pi, 0)$ and $(0.4\pi, 0)$, respectively, in addition to that at $(\pi, 0)$. While former peak structures 
result from the intrapocket nesting, the latter one arises due to the interpocket nesting. Moreover in the former case, a small elongation of the electron pocket along $k_x$ [$k_y$] and it's  predominant 
$d_{yz}$ [$d_{xz}$] orbital character at ($\pi, 0$) [($0, \pi$)] results into the difference of peak positions. We also note that $\chi_{2222}(\q)$ displays stronger peak at $(\pi, 0)$ due to the relatively
good interpocket nesting between the electron pocket at ($\pi, 0$) and hole pocket at (0, 0) than between the electron pocket at ($0, \pi$) and hole pocket at ($\pi, \pi$).  On the other hand, $\chi_{1122}(\q)$ has a broad peak at $(\pi, 0)$ with the significant contribution coming from the interpocket nesting between the electron and hole pockets, and therefore it 
contributes significantly to the SDW instability with ordering wavevector $(\pi, 0)$ in this model. However, one bubble orbital susceptibility shows maximum near (0, 0) in contrast 
with the spin susceptibility. This follows from the subtraction of the peak structure of 
$\chi_{1122}(\q)$ from $\chi_{1111}(\q)$ and $\chi_{2222}(\q)$ according to Eq. \ref{orb}. 
  
Fig. 1(d) shows the RPA orbital susceptibility diverging near $\q \approx (0, 0)$ for $U = 3.0$ and $\lambda = 0.3U$, which implies the existence
of ferro orbital-ordering instability in this model. Here, we recall that the $(\pi, 0)$-SDW state is stabilized 
in the absence of any orbital-phonon coupling as reported in earlier studies.\cite{raghu,ghosh,brydon} We further note that $\lambda$ as small as $\approx 0.075W$ with $U$ $\approx 0.25W$
can lead to orbital-ordering instability, where $W$ is 
estimated to be $\approx 4eV$ from the band-structure calculation so that $\lambda \approx 0.3eV$. Therefore, $\tilde{\lambda} \approx 0.12$ as the density of state at the Fermi level is $\rho$ 
$\approx 0.4/eV$.
  
The ferro-type orbital-ordering instability remains largely unaffected for very small electron or hole doping as 
the Fermi surface does not display any sharp change in it's shape, especially in the electron-doped regime. However, the electron pockets disappear on hole doping near $x_h \approx 0.18$ so 
that there is a dramatic change in the nesting condition. Consequently, an antiferro orbital-ordering instability appears due to the interpocket nesting 
between the hole pockets at (0, 0) and at ($\pi, \pi$), which is also reflected in the RPA orbital susceptibility as
shown in Fig. 1(e). The critical $\tilde{\lambda} \approx 0.18$ is comparatively larger than that for the undoped case 
owing to the relatively poor nesting. 

\subsection{Three-orbital model of Daghofer \textit{et al}.}
The Fermi surface in the three-orbital model of Daghofer \textit{et al}., whose basis also includes $d_{xy}$ orbital
in addition to $d_{xz}$ and $d_{yz}$ orbitals, is shown (Fig. \ref{daghofer}(a)) for the chemical 
potential $\mu = 0.21$ ($n \approx$ 4).
Several limitations of the two-orbital model such as a large hole pocket at ($\pi, \pi$), the absence of
an additional hole pocket at (0, 0), and the contribution of $d_{xy}$ orbital to the 
Fermi surface are remedied. Also, the electron pockets at ($\pi$, 0) is elliptical in accordance with the ARPES measurement.

Fig. \ref{daghofer}(b) shows the components of susceptibility for $n \approx$ 4. Both $\chi_{1111}(\q)$ and $\chi_{2222}(\q)$ have peaks 
close to ($0.4\pi, 0$) and ($0.6\pi, 0$) arising due to the intrapocket nesting. The peak for $\chi_{1111}(\q)$ results from the nesting along the 
minor axis of the elliptical electron pocket dominated by $d_{xz}$ orbital at ($0, \pi$), while the peak for $\chi_{2222}(\q)$ follows from
the nesting in the outer hole pocket dominated by 
$d_{yz}$ along $k_x$. In addition, $\chi_{2222}(\q)$ has a peak near ($\pi, 0$) for the reason described in the previous section. 
However, $\chi_{1111}(\q)$ does not show any peak structure close to ($\pi, 0$) due to the absence of the hole pocket at ($\pi, \pi$) unlike the two-orbital model.
Another difference from two-orbital model is that $\chi_{1122}(\q)$ is negligibly small. For these reasons,
the bare orbital susceptibility exhibits peak along a ring-shaped structure centered around $\q = {\bf 0}$ as shown in Fig. \ref{daghofer}(c).
  
The RPA orbital susceptibility displays divergence behavior near $ {\q_1} \approx (0.4\pi, 0)$ and $(0, 0.4\pi)$ for $U = 0.6eV$ and $\lambda = 0.27U$(Fig. \ref{daghofer}(d)). The critical dimensionless electron-phonon coupling parameter
$\tilde{\lambda} \approx$ 0.33 is almost three times larger than that in the model of Raghu \textit{et al}. because of the large density
of state at the Fermi level $\rho \approx 2/eV$. Here, orbital-lattice 
coupling leads to an incommensurate orbital-ordering instability instead of ferro orbital-ordering instability, which results from a relatively
strong intrapocket nesting along the minor axis of the elliptical electron pocket. This nesting can also play a significant role in stabilizing the magnetic 
order in the absence of coupling to the lattice degree of freedom. It is worthwhile to mention that this model exhibits 
ferromagnetic order instead of the $(\pi, 0)$-SDW state for small intraorbital Coulomb interaction.\cite{daghofer} Further, the hole pockets become smaller and coincident on electron doping while the electron pocket
remains largely unaffected, which leads to the shifting of peak structure towards $\q = {\bf 0}$ because of the intrapocket nesting of 
the hole pockets as shown in Fig. \ref{daghofer}(e). While the magnitude of incommensurate nesting vector increases on hole doping. 

\subsection{Five-orbital model of Graser \textit{et al}.}
Finally, we consider the five-orbital model of Graser \textit{et al}.. The size of the electron and hole pockets are now relatively smaller in comparison to
the three-orbital model as shown Fig. \ref{graser}(a). The ellipticity of the electron pocket is also reduced. 
Another important difference being the the good nesting between the electron and hole pockets, 
which gives rise to $(\pi, 0)$ SDW in the undoped compound.\cite{graser} These features are similar to the other 
five-orbital models.\cite{kuroki}

The components of the susceptibility are shown in Fig. \ref{graser}(b). Incommensurate peak structures for $\chi_{1111}(\q)$ and $\chi_{2222}(\q)$ near (0, 0) are weaker whereas the peak of $\chi_{2222}(\q)$ 
close to ($\pi, 0$) is strong due to the good interpocket nesting in comparison to the three-orbital model. The shape of the outer hole pockets is intermediate between a circle and a square, therefore there 
exists an intrapocket nesting along $k_x$, which leads to the peak in $\chi_{2222}(\q)$ near $(0.36\pi, 0)$. Similarly, the peak structure corresponding to the small momentum in $\chi_{1111}(\q)$ results from the 
intrapocket nesting of the inner hole pocket. Therefore, the bare orbital susceptibility exhibits peak near $(0, 0)$
along a ring-shaped structure as well as near ($\pi, 0$) (Fig. \ref{graser}(c)).  

We note two differences in RPA-level orbital susceptibility from the three-orbital model. The divergence of orbital susceptibility near $(0.36\pi, 0)$ as shown in Fig. \ref{graser}(d) implies that the intrapocket nesting in the outer
hole pocket is important for the orbital ordering. Secondly, the momentum corresponding to the instability dramatically gets relocated near ($\pi, 0$) due to the improved interpocket nesting (Fig. \ref{graser}(e)). However,
the critical electron-phonon coupling parameter is ${\lambda} \approx 0.6$, so that $\tilde{\lambda} \approx 0.36$ as $\rho = 0.6/eV$, which is similar in magnitude to 
that in the case of three-orbital model.  

\section{Conclusions and Discussions}
In conclusion, we have investigated orbital-ordering instability in the models of pnictides resulting from intrapocket nesting in the presence of 
orbital-lattice coupling. Such instabilities are sensitive to the shape of the Fermi surface, and therefore are dependent on doping as well as on the model. In the undoped case, two-orbital model of 
Raghu \textit{et al}. has an instability toward ferro orbital order whereas  three-orbital model of Daghofer \textit{et al}. displays an instability towards 
an incommensurate order with small momentum. In the latter, the instability arises due to relatively strong nesting along the minor axes of the elliptical electron pocket. Such instability may 
also be exhibited by the model of Calder\'{o}n \textit{et al}.\cite{calderon} because of a large major to minor axis ratio of the elliptical electron pocket. In the five-orbital model of Graser \textit{et al}., 
the instability results from the intrapocket nesting of the hole pocket. The dimensionless electron-phonon coupling parameters required to induce such instability in the models of Raghu \textit{et al}., Daghofer \textit{et al}., and 
Graser \textit{et al}., are $\approx$ 
0.12, 0.36, and  0.36, respectively. Therefore, the value of coupling parameter is smaller in the two-orbital model than that obtained from the first-principle calculation ($\tilde{\lambda} \approx 0.21$),\cite{boeri} 
while slightly larger in the three- and five-orbital models. 

\section* {Acknowledgements} 
The author is indebted to Tetsuya Takimoto for useful discussions.
This work is supported by Basic Science Program through the National Research Foundation of Korea (NRF) funded by the Ministry of Education (NRF-2012R1A1A2008559). 

\end{document}